\documentclass[a4paper,11pt]{article}
\usepackage{pos}
\usepackage{float}
\usepackage{flexisym}
\usepackage{lmodern}

\title{Illuminating long-lived dark vector bosons via exotic Higgs decays at $\sqrt{s} = 13\,{\text {TeV}}$}

\author*[a]{Tamer Elkafrawy}
\author[a]{Marcus Hohlmann}
\author[b]{Teruki Kamon}
\author[c]{Paul Padley}
\author[b]{Hyunyong Kim}
\author[a]{Mehdi Rahmani,}
\author[c]{Sven Dildick}

\affiliation[a]{Florida Institute of Technology,\\
	Melbourne, Florida USA}

\affiliation[b]{Texas A\&M University,\\
	College Station, Texas USA}

\affiliation[c]{Rice University,\\
	Houston, Texas USA}

\emailAdd{telkafrawy@fit.edu}
\emailAdd{tamer.elkafrawy@cern.ch}
\emailAdd{taelkafr@fnal.gov}

\abstract{The possibility of producing a measurable long-lived dark $Z$ boson, that is assumed to be produced either via its kinetic mixing with the hypercharge gauge boson or via mixing of the observed 125-GeV Higgs boson with the dark Higgs boson, called Higgs mixing, in Run 2 of the Large Hadron Collider (LHC) is investigated. Displaced dimuons in the final state are considered where each of the $Z$ and the dark $Z$ bosons decays directly to a dimuon. The Higgs production cross sections via gluon-gluon fusion at 13 TeV calculated to a combination of next-to-next-to-next-to-leading order with QCD corrections (N$^{3}$LO QCD) and next-to-leading order with electroweak corrections (NLO EW) from the literature are used, while the branching fractions are calculated to NLO by using Monte Carlo simulation in the framework of {\textsc{MadGraph5}}\_aMC@NLO and compared to the available analytical calculations to leading order (LO). Sensitivities of the LHC in Run 2 to such searches are discussed.}

\begin{document}
	\maketitle
	\section{Introduction}
	Exotic Higgs decays involve new light states beyond the Standard Model (BSM) and are best searched at the LHC. In this context, the observed 125-GeV Higgs boson $h$ is assumed to be responsible for breaking the electroweak symmetry and to decay to new particles such as dark Higgs boson and dark $Z$ boson, usually referred to as $h_D$ (or $s$) and $Z_D$, respectively. The only possible interaction of $Z_D$ with the Standard Model (SM) sector is through its kinetic mixing (KM) with $Z$ boson \cite{Galison1984,Holdom1986,Dienes1997}, while if the Higgs mixing (HM) exists, $h_D$ will have a renormalizable coupling to $h$. The high luminosities achieved by the LHC offer a promising insight into the search for hidden sectors through these two portals. The search for $Z_D$ has been mostly performed for $m_{Z_D}<10$~GeV, while several efforts were recently devoted to higher ranges of $m_{Z_D}$ \cite{Curtin2014,Davoudiasl2013,Gopalakrishna2008,Chang2014,Falkowski2014}. Heavier $Z_D$ could explain the $\sim3.6\sigma$ discrepancy between the measured and SM value of the muon anomalous magnetic moment \cite{Pospelov2009,Bennett2006,Abi2021} as well as various dark matter (DM)-related anomalies via new DM-$Z_D$ interactions \cite{Arkani-Hamed2009,Pospelov2009a,Finkbeiner2007,Fayet2004}. In this investigation, decays are assumed to proceed through on-shell $Z_D$'s (i.e., $m_{Z_D}<m_h/2$ in the case of $h\rightarrow{Z_DZ_D}$ and $m_{Z_D}<m_h-m_{Z}$ in the case of $h\rightarrow{ZZ_D}$) for which cross sections are enhanced as compared to the off-shell $Z_D$'s for which cross sections are suppressed by the second power of the kinetic mixing parameter $\epsilon^2$ \cite{Curtin2015}. Kinematic acceptances and efficiencies of 100\% are assumed for all NLO simulated observables, which are scanned in two-dimensional scans over the relevant free parameters, enabling us to come up with constraints on such free parameters, which are the kinetic and Higgs mixing parameters as well as the acquired mass by $Z_D$. Samples are generated by using Monte Carlo (MC) simulation in the framework of {\textsc{MadGraph5}}\_aMC@NLO v2.7.0. Feynman diagrams of the two decays are given in Fig.~\hyperlink{Fig.1}{1}.
	
	\begin{figure}[H]
		\begin{center}
			\includegraphics[width=0.85\textwidth]{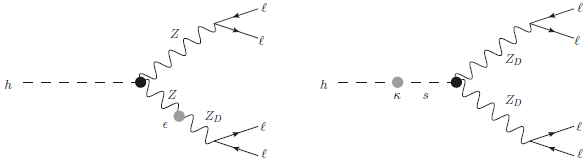}
			\caption{Exotic Higgs decays to a final state of four leptons with an intermediate $Z_D$ produced through Kinetic mixing via the hypercharge portal (left) or Higgs mixing via the Higgs portal (right) \cite{Curtin2014}. The dashed line indicates that mixing of the SM bosons with their dark sector counterparts is occurring in the decays.
			}
		\end{center}
		\hypertarget{Fig.1}{}
	\end{figure}
	
	\section{Exotic Higgs decay widths}
	The exotic Higgs decay $h\rightarrow{Z_DZ_D}\rightarrow{2\mu^{+}2\mu^{-}}$ is induced if the HM dominates irrespective of the KM size, while the exotic Higgs decay $h\rightarrow{ZZ_D}\rightarrow{2\mu^{+}2\mu^{-}}$ is induced if the KM dominates irrespective of the HM size. The exotic decay width $\Gamma(h\rightarrow{Z_DZ_D})$ to LO in the HM parameter $\kappa$, and in turn the corresponding branching fraction and total cross section $\sigma_{total}$, is tremendously impacted by $\kappa$, less impacted by $m_{Z_D}$, and much less impacted by $m_{h_D}$ as given in Ref.~\cite{Curtin2014} by
	
	\begin{equation}
		\centering
		\Gamma(h\rightarrow{Z_DZ_D}) = \frac{\kappa^2}{32\pi}\frac{\upsilon^2}{m_h}\sqrt{1-\frac{4m_{Z_D}^2}{m_h^2}}\frac{(m_h^2+2m_{Z_D}^2)^2-8(m_h^2-m_{Z_D}^2)m_{Z_D}^2}{(m_h^2-m_{h_D}^2)^2},
		\label{1}
	\end{equation}
	where $\upsilon=246$~GeV is the SM Higgs vacuum expectation value (vev). The partial decay width $\Gamma(h\rightarrow{ZZ_D})$ to LO in $m_{Z_D}^2/m_Z^2$ is highly impacted by $\epsilon$ and less impacted by $m_{Z_D}$ as given in Ref.~\cite{Curtin2014} by
	
	\begin{equation}
		\centering
		\Gamma(h\rightarrow{ZZ_D}) = \frac{\epsilon^2tan^2\theta_w}{16\pi}\frac{m_{Z_D}^2(m_h^2-m_Z^2)^3}{m_h^3m_Z^2\upsilon^2},
		\label{2}
	\end{equation}
	where $\theta_w$ is the Weinberg mixing angle that is measured as $\sim28.75^\circ$ by LHCb \cite{Aaij2015}. The upper two panels of Fig.~\hyperlink{Fig.2}{2} are generated to compare the analytical calculation (upper right) as given by Eq.~\eqref{1} to the simulation (upper left) of $\Gamma(h\rightarrow{Z_DZ_D})$ in a scan over the $\kappa$-$m_{Z_D}$ plane, while the lower two panels of this figure are generated for the same sake of comparison for $\Gamma(h\rightarrow{ZZ_D})$.
	
	\begin{figure}[H]
		\centering
		\includegraphics[width=0.445\textwidth]{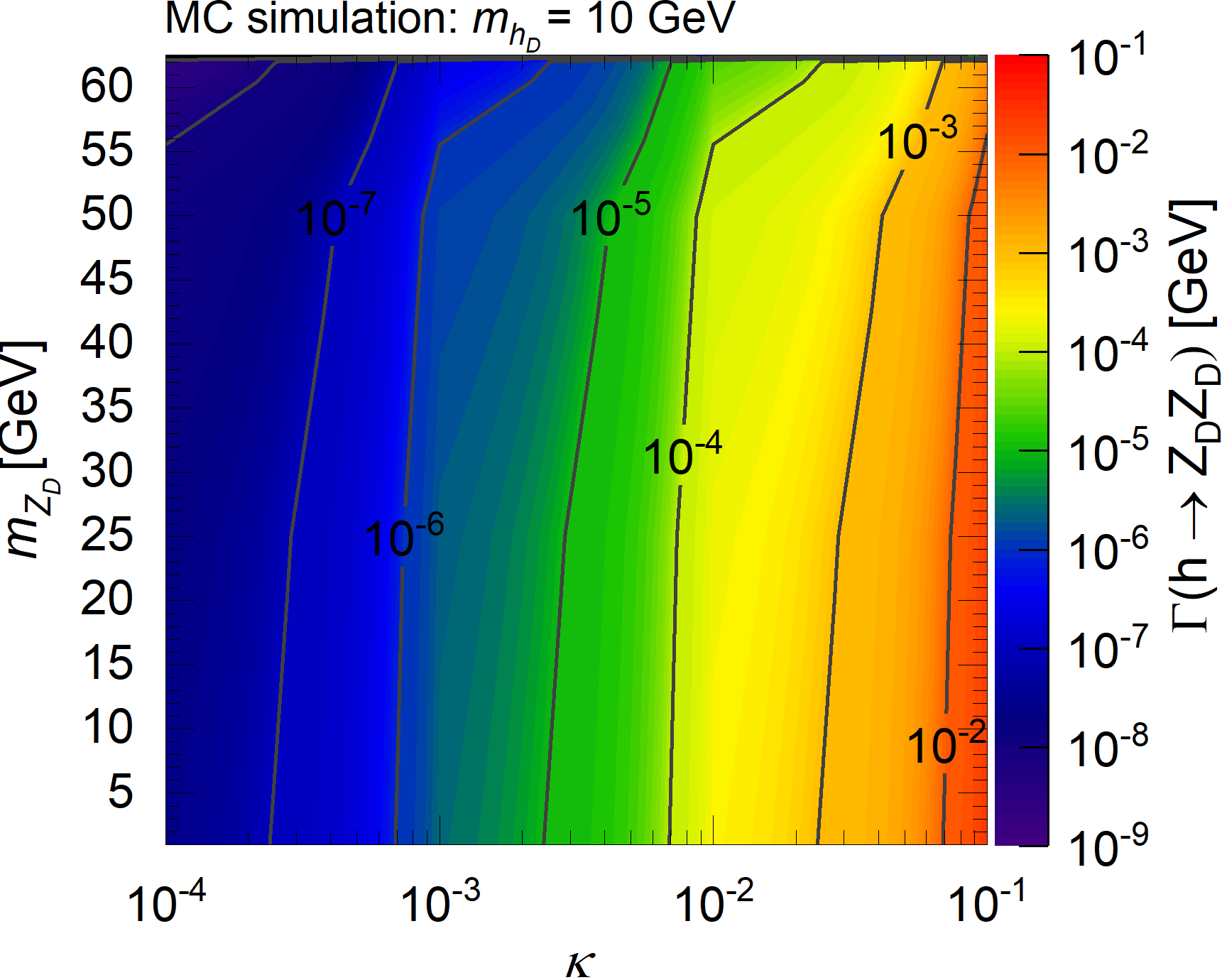}
		\includegraphics[width=0.445\textwidth]{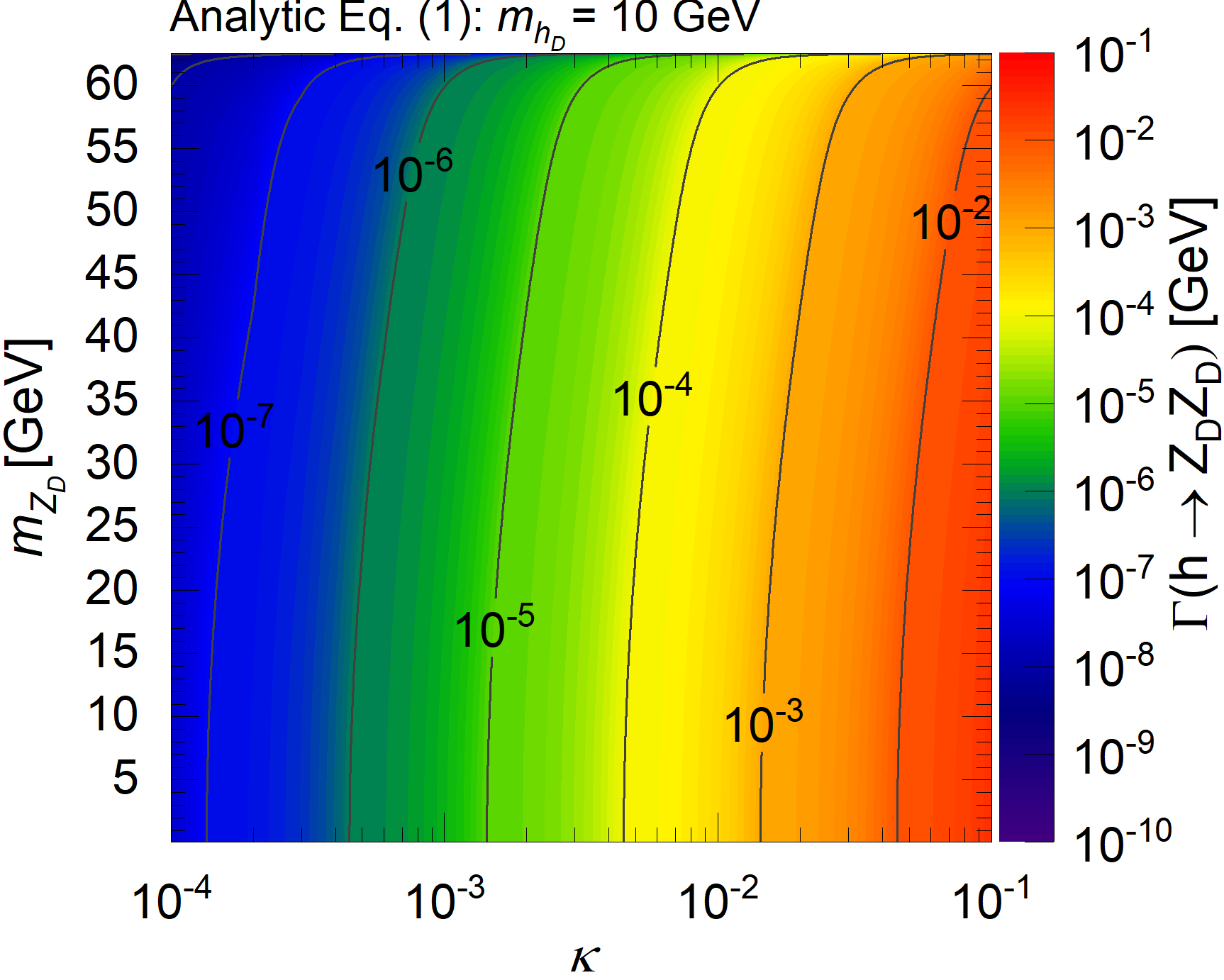}
		\includegraphics[width=0.445\textwidth]{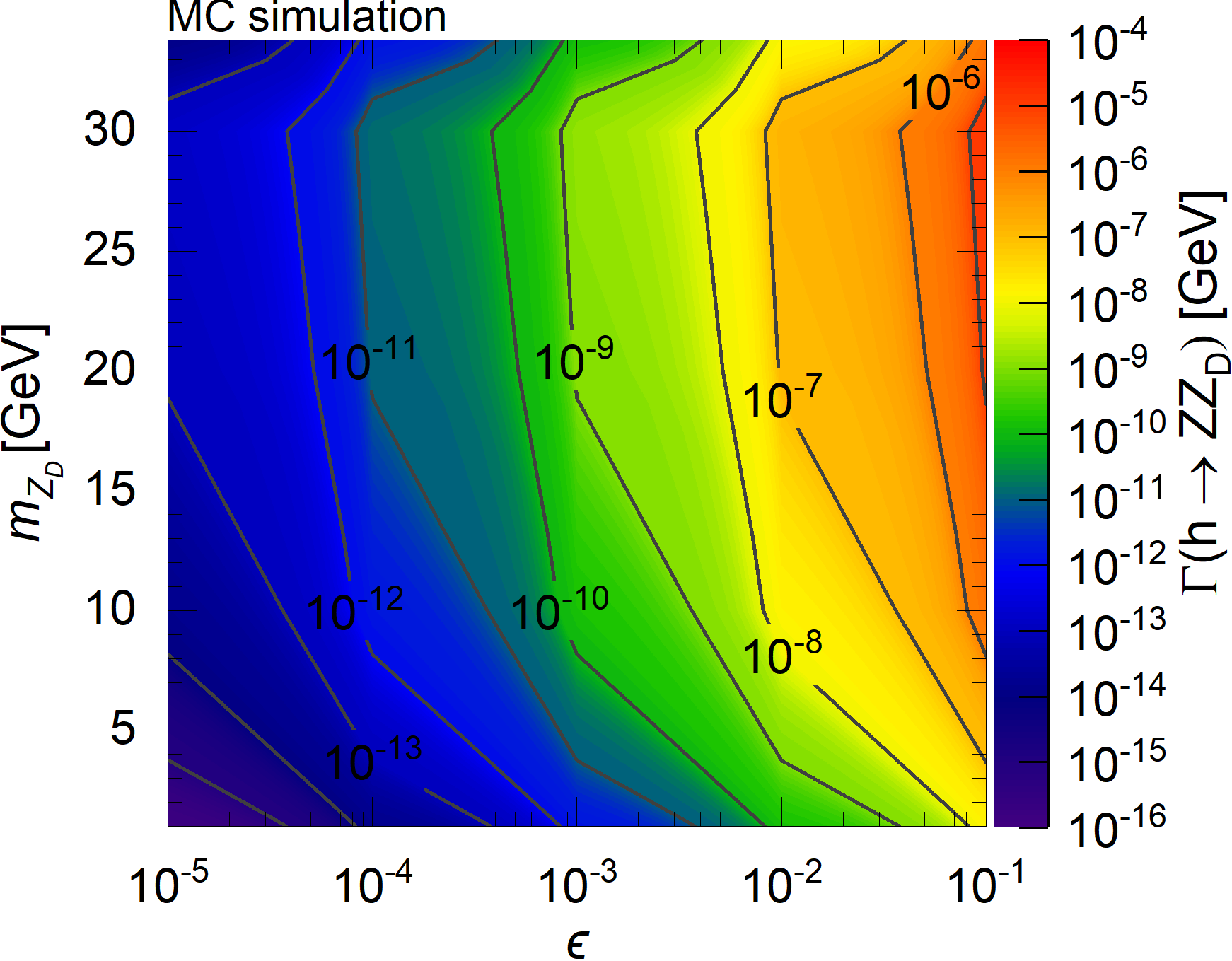}
		\includegraphics[width=0.445\textwidth]{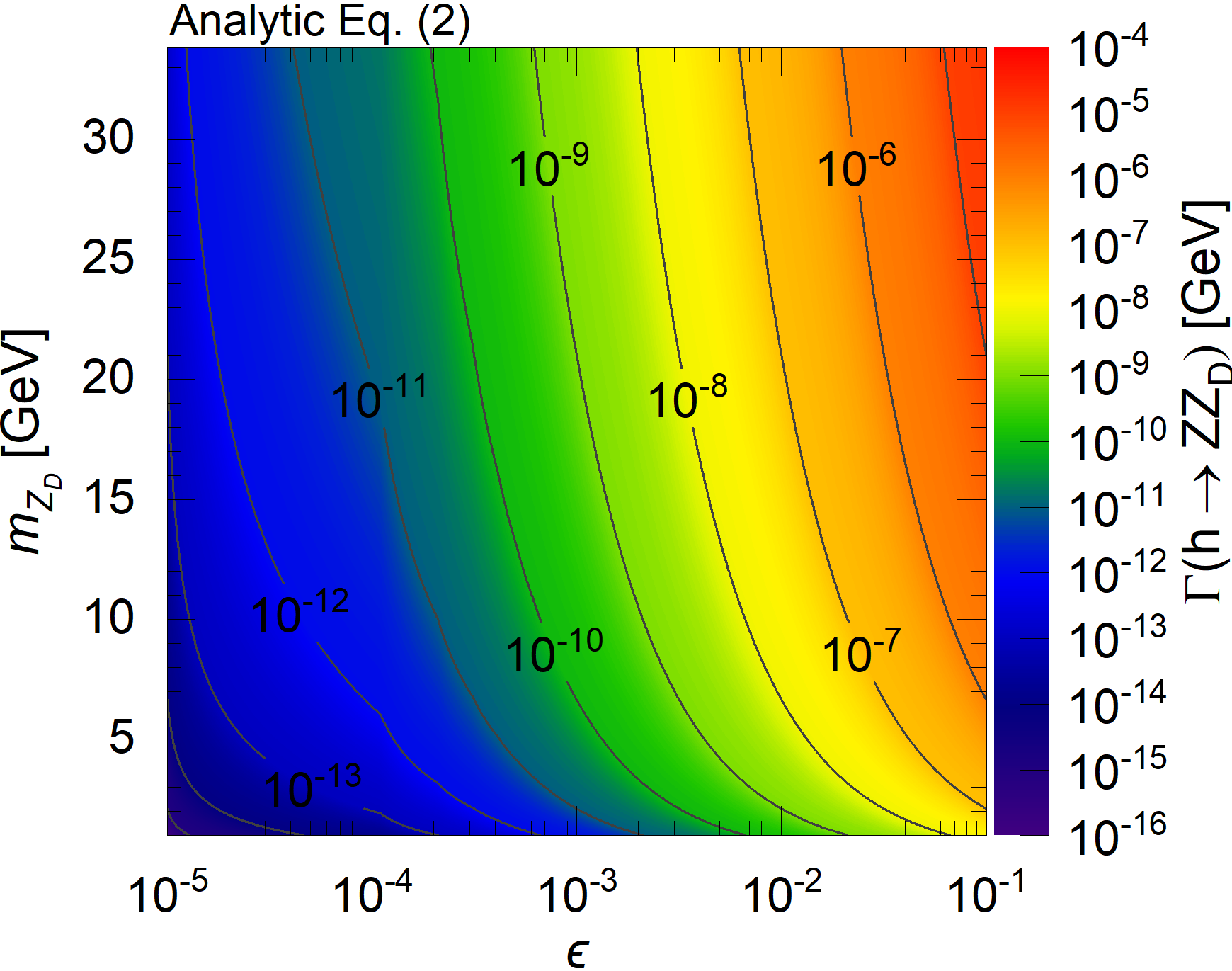}
		\caption{MC simulation (upper left) against an analytical calculation (upper right) from Eq.~\eqref{1} of $\Gamma(h~\rightarrow~{Z_DZ_D})$ in a scan over the $\kappa$-$m_{Z_D}$ plane for $m_{h_D}=10$~GeV as well as MC simulation (lower left) against an analytical calculation (lower right) from Eq.~\eqref{2} of $\Gamma(h\rightarrow{ZZ_D})$ in a scan over the $\epsilon$-$m_{Z_D}$ plane.}
		\hypertarget{Fig.2}{}
	\end{figure}
	
	\section{Branching fractions of exotic Higgs decays}
	The branching fraction of $Z_D$ to light leptons is appreciable for any $m_{Z_D}<m_h/2$ \cite{Curtin2014}. Owing to the caption of Fig.~\hyperlink{Fig.3}{3}, the far left panel shows an excellent agreement between the analytical calculation (dashed orange), derived from Eq.~\eqref{1}, and the MC simulation (solid black) of $B(h\rightarrow{Z_DZ_D})$, which varies directly with $\kappa^2$ \cite{Curtin2014,Curtin2015} until the HM becomes too small to handle the decay, which then proceeds through KM and varies directly with $\epsilon^4$ \cite{Curtin2015}. The later behavior disappears from the LO-based analytic curve (dashed orange) but still exists in the NLO-based simulated curve (solid black). In the middle left panel of this figure, $B(h\rightarrow{ZZ_D})$ is shown to vary directly with $\epsilon^2$ \cite{Curtin2014,Curtin2015}, while the scans over $m_{Z_D}$ go as low as 1 GeV. The four panels of this figure show that $B(Z\rightarrow\mu^{+}\mu^{-})$ and $B(Z_D\rightarrow\mu^{+}\mu^{-})$ are unchanged regardless of the scan.
	
	\begin{figure}[H]
		\centering
		\includegraphics[width=0.43\textwidth]{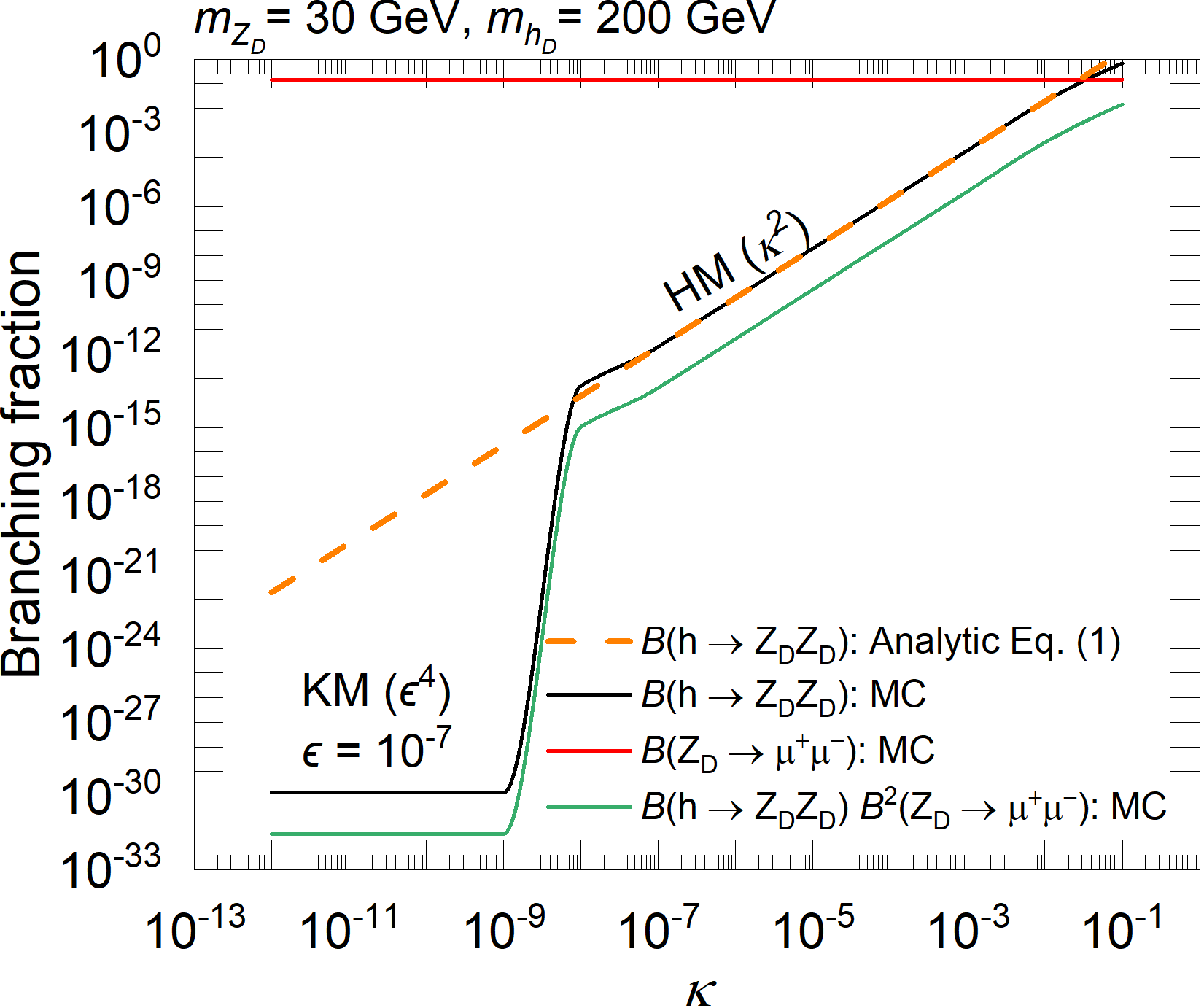}
		\includegraphics[width=0.43\textwidth]{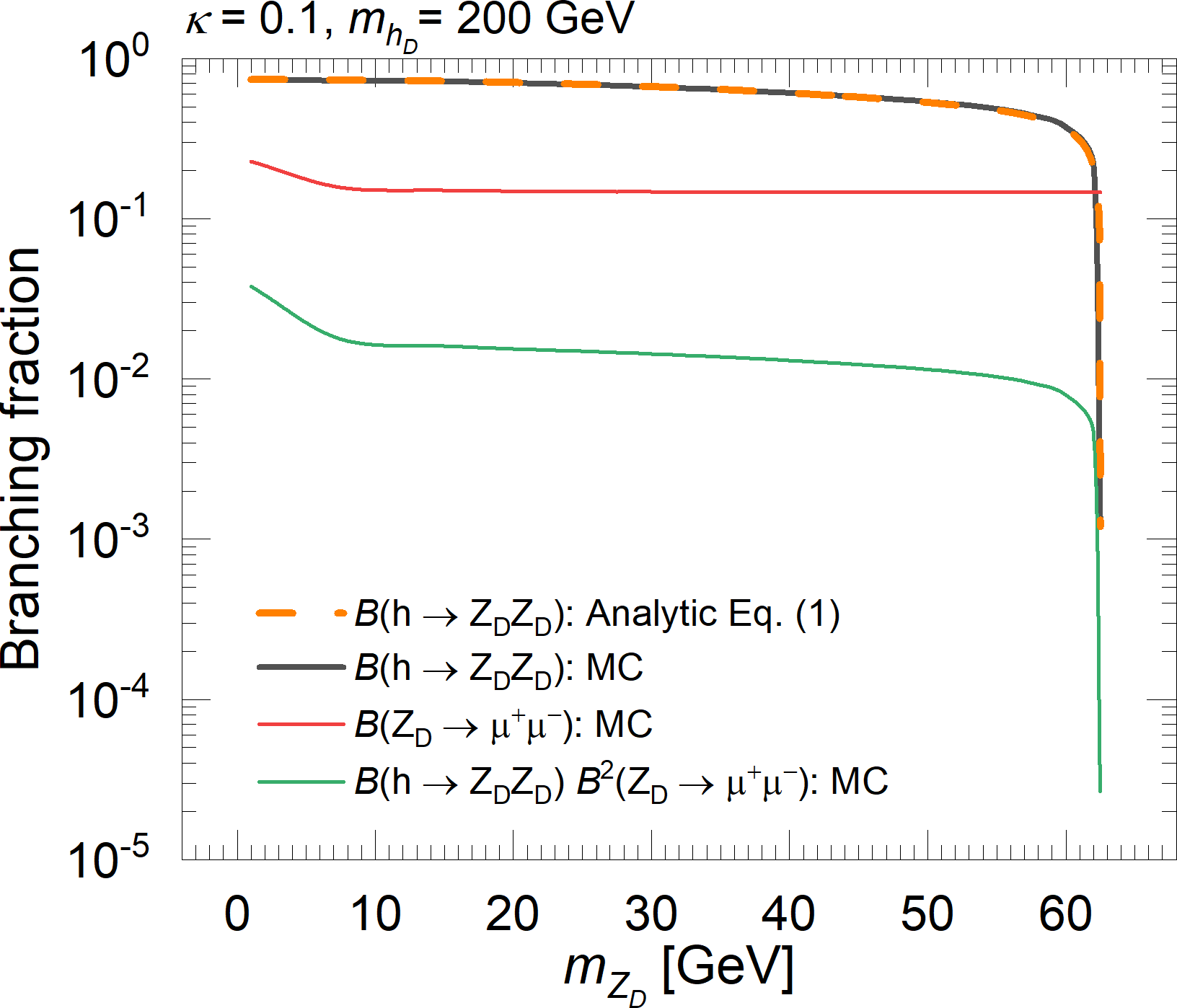}
		\includegraphics[width=0.43\textwidth]{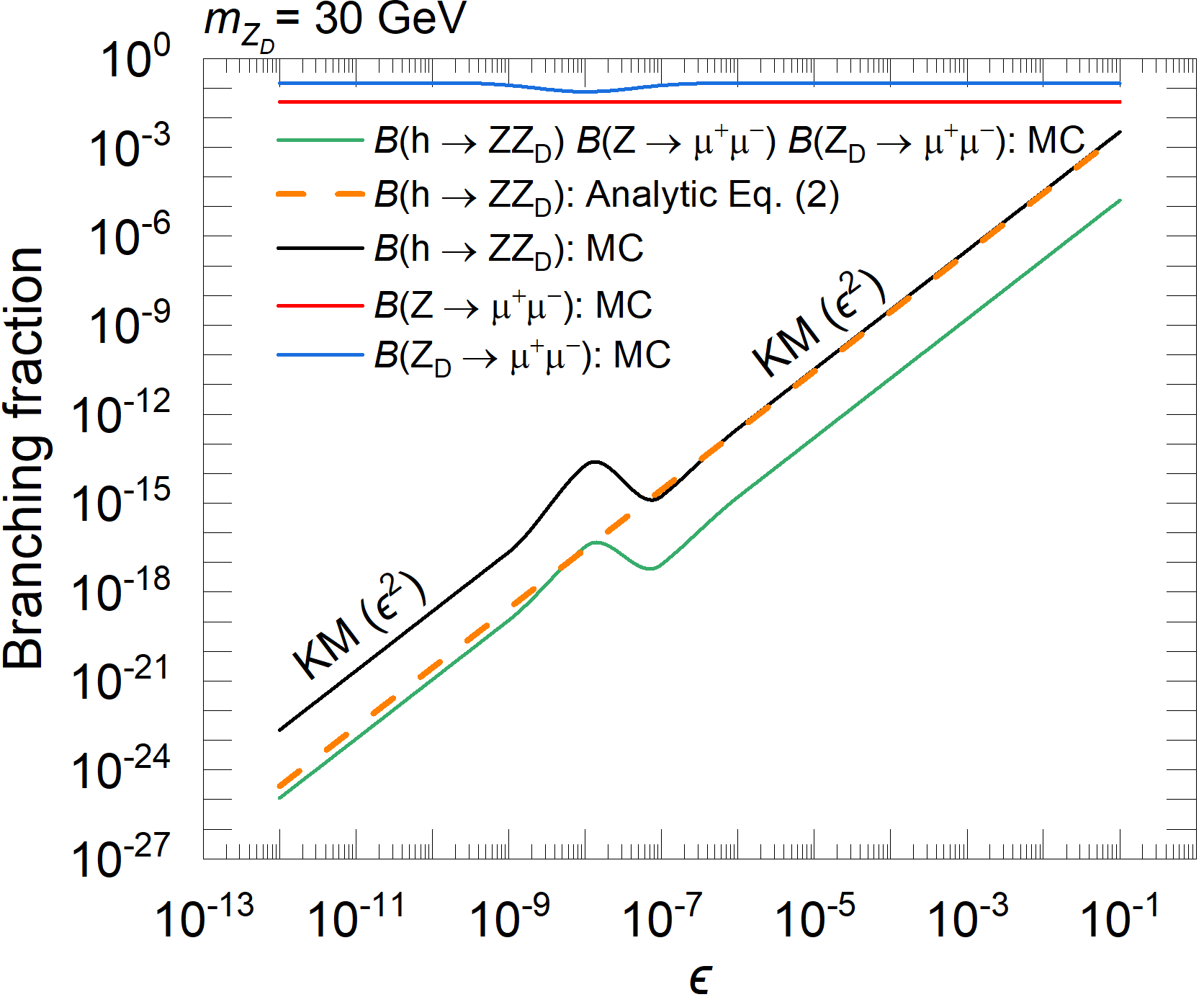}
		\includegraphics[width=0.43\textwidth]{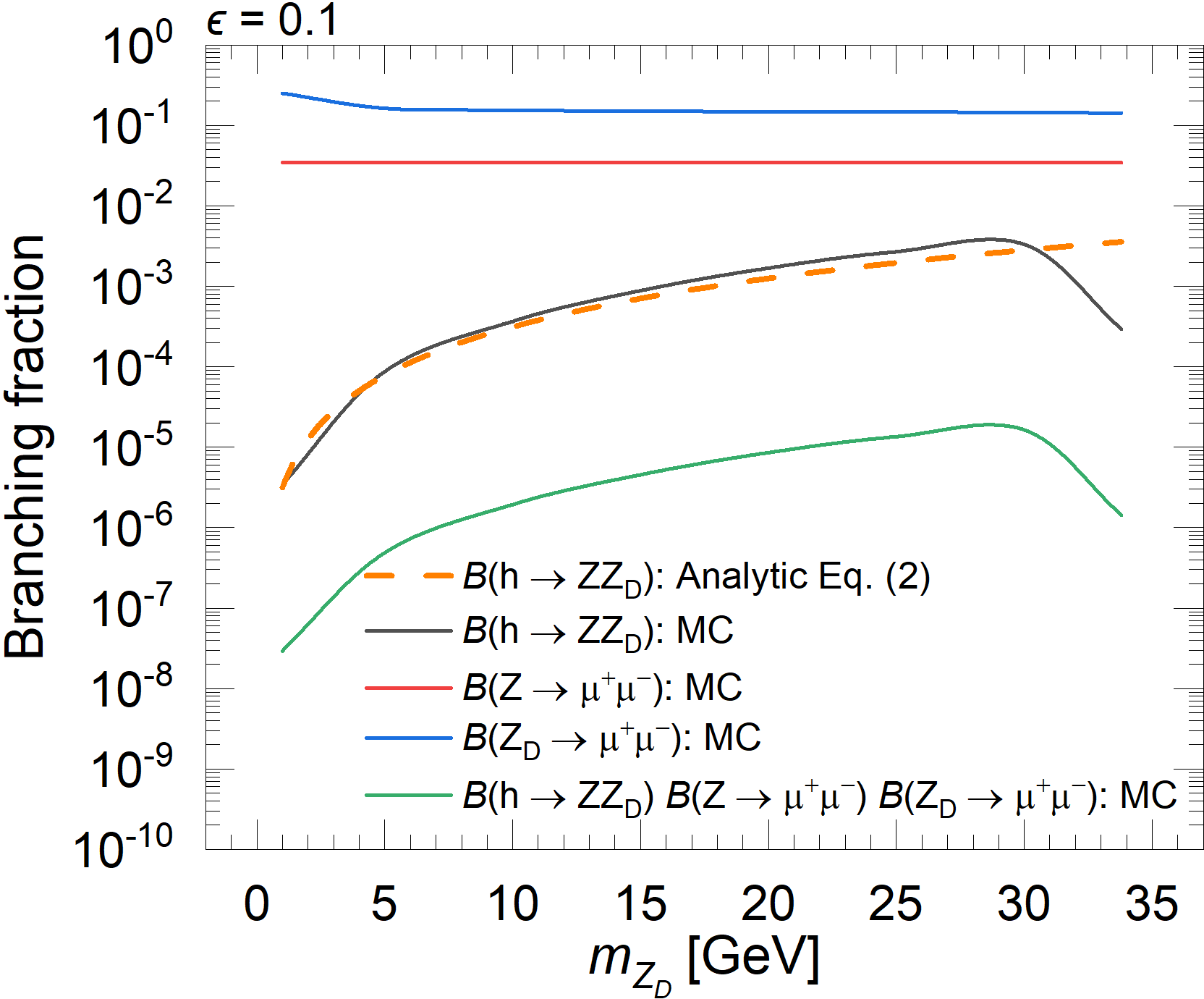}
		\caption{MC simulation of $B(h\rightarrow{Z_DZ_D})$, $B(Z_D\rightarrow{\mu^{+}\mu^{-}})$, and the product of $B(h\rightarrow{Z_DZ_D})$ and $B^2(Z_D\rightarrow{\mu^{+}\mu^{-}})$ as well as an analytical calculation of $B(h\rightarrow{Z_DZ_D})$ from Eq.~\eqref{1} in a scan over $\kappa$ (far left) and $m_{Z_D}$ (middle left), and also MC simulation of $B(h\rightarrow{ZZ_D})$, $B(Z\rightarrow{\mu^{+}\mu^{-}})$, $B(Z_D\rightarrow{\mu^{+}\mu^{-}})$, and their product as well as an analytical calculation of $B(h\rightarrow{ZZ_D})$ from Eq.~\eqref{2} in a scan over $\epsilon$ (middle right) and $m_{Z_D}$ (far right).}
		\hypertarget{Fig.3}{}
	\end{figure}
	
	\section{Sensitivity of the LHC in Run 2 to the searches for long-lived dark $Z$ bosons}
	
	The SM Higgs is considered to be produced through the production channel of gluon-gluon fusion (ggF) for which the production cross section of $\sim48$~pb, calculated to a combination of next-to-next-to-next-to-leading order with QCD corrections (N$^{3}$LO QCD) and next-to-leading order with electroweak corrections (NLO EW) from the literature \cite{Florian2017}, is used, while the branching fractions are calculated to NLO by using MC simulation in the framework of {\textsc{MadGraph5}}\_aMC@NLO v2.7.0. The $Z_D$ decay length ($c\tau_{Z_D}$) is controlled by $\epsilon$ and $m_{Z_D}$ only and inversely proportional to $\epsilon^2$ regardless of the exotic decay mode. The value of $\epsilon=10^{-7}$ is selected to generate the left panel of Fig.~\hyperlink{Fig.4}{4}. This value of $\epsilon$ along with a range of $m_{Z_D}$ of $1-62.49999$~GeV corresponds to a $c\tau_{Z_D}$ range of $10-2000$~mm, that is measurable within the geometrical size of the Compact Muon Solenoid (CMS) detector. The value of 0.073~fb is taken as the smallest $\sigma_{total}$ to which the LHC is sensitive based on 10 events at least to be measured in Run 2 of the LHC with its full integrated luminosity of 137~fb$^{-1}$.
	\begin{figure}[H]
		\begin{center}
			\includegraphics[width=0.4\textwidth]{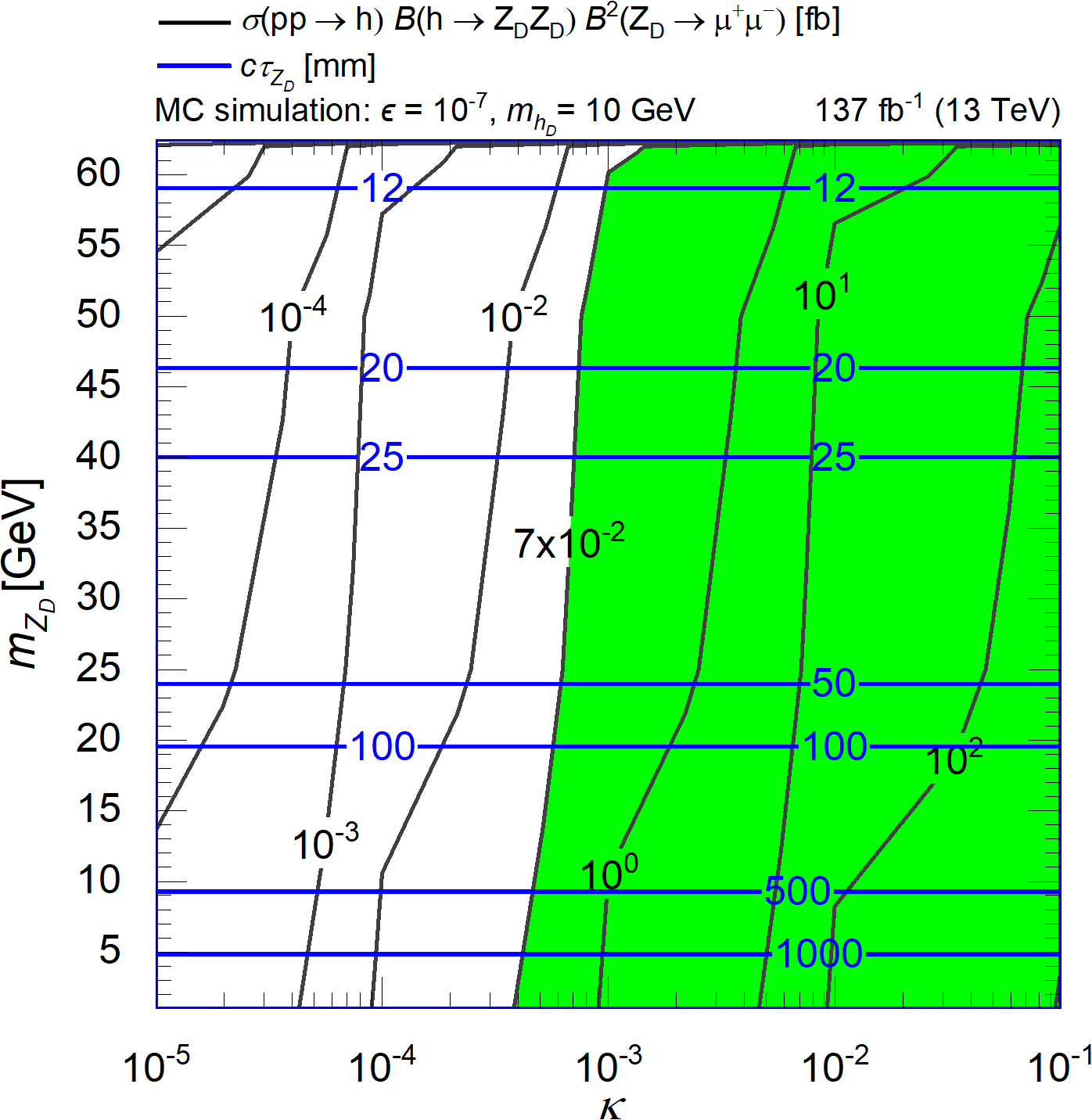}
			\includegraphics[width=0.4\textwidth]{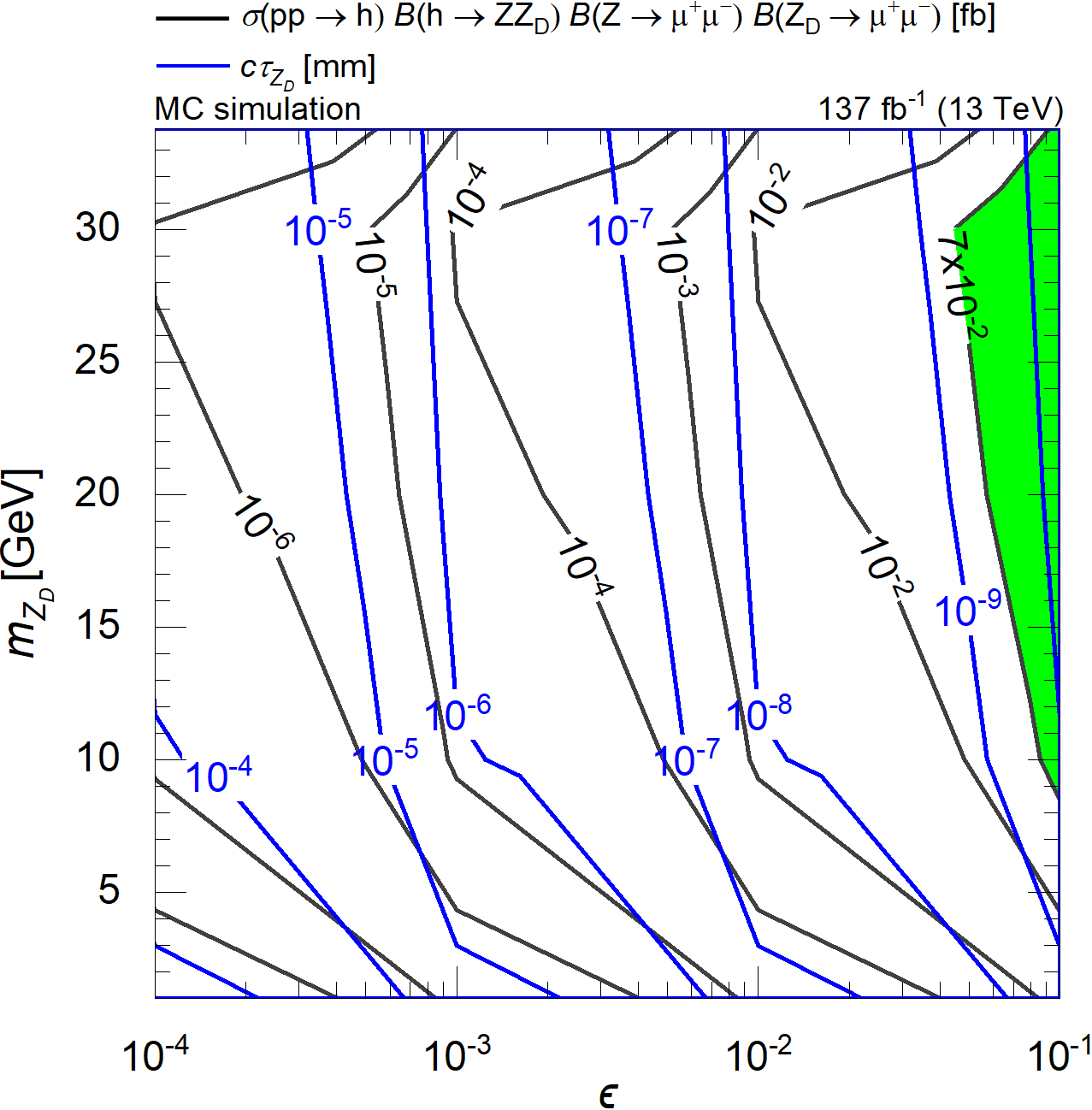}
			\caption{MC simulation showing the contour lines of $\sigma_{total}$ (black) and $c\tau_{Z_D}$ (blue) for the exotic Higgs decay mode $h\rightarrow{Z_DZ_D}\rightarrow{2\mu^{+}2\mu^{-}}$ (left) for $m_{h_D}=10$~GeV and the exotic Higgs decay mode $h\rightarrow{ZZ_D}\rightarrow{2\mu^{+}2\mu^{-}}$ (right) in a scan over the $\kappa$-$m_{Z_D}$ plane and the $\epsilon$-$m_{Z_D}$ plane, respectively, for Run 2 of the LHC for which sensitivity regions are shaded in green.}
		\end{center}
		\hypertarget{Fig.4}{}
	\end{figure}
	
	\section{Conclusion}
	The LHC in Run 2 is sensitive to the searches for dark $Z$ bosons via $h\rightarrow~{Z_DZ_D}\rightarrow{2\mu^{+}2\mu^{-}}$ down to $\kappa=4\times10^{-4}$ irrespective of the mass acquired by the dark $Z$ boson. A long-lived dark $Z$ boson is likely to be observed at the LHC with a decay length in the range of $10-2000$~mm if $\epsilon\sim10^{-7}$ is selected for the kinetic mixing parameter for a range of $1-62.49999$~GeV of the dark $Z$ mass with the decay length and the mass being inversely proportional to each other. A larger kinetic mixing leads to the production of a dark $Z$ boson with a shorter lifetime, and vice versa. The dark $Z$ boson is less likely to be measured in Runs 2 and 3 of the LHC via $h\rightarrow{ZZ_D}\rightarrow{2\mu^{+}2\mu^{-}}$. The LHC in Run 2 is sensitive to this decay mode down to $\epsilon=4.5\times10^{-2}$, which restricts the produced dark $Z$ boson to being prompt, while the acquired mass by the dark $Z$ boson has to fall in the range of $9-33.8$~GeV if $\epsilon$ falls in the range of $4.5\times10^{-2}-10^{-1}$ and in the range of $<9$~GeV if $\epsilon$ falls in the range of $>10^{-1}$.


\begin{thebibliography}{99}
		
		\bibitem{Galison1984}
		P. Galison and A. Manohar, \emph{Two $Z$'s or not two $Z$'s?}, \href{https://doi.org/10.1016/0370-2693(84)91161-4} {\emph{Phys. Lett. B} {\bf136} (1984) 279}.
		
		\bibitem{Holdom1986}
		B. Holdom, \emph{Two U(1)'s and $\epsilon$ charge shifts}, \href{https://doi.org/10.1016/0370-2693(86)91377-8} {\emph{Phys. Lett. B} {\bf166} (1986) 196}.
		
		\bibitem{Dienes1997}
		K.R. Dienes, C.F. Kolda, and J. March-Russell, \emph{Kinetic mixing and the supersymmetric gauge hierarchy}, \href{https://doi.org/10.1016/S0550-3213(97)00173-9} {\emph{Nucl. Phys. B} {\bf492} (1997) 104} [arXiv:hep-ph/9610479].
		
		\bibitem{Curtin2014}
		D. Curtin \emph{et al.}, \emph{Exotic decays of the 125 GeV Higgs boson}, \href{https://doi.org/10.1103/PhysRevD.90.075004} {\emph{Phys. Rev. D} {\bf90} (2014) 075004} [arXiv:1312.4992].
		
		\bibitem{Davoudiasl2013}
		H. Davoudiasl, H.-S. Lee, I. Lewis, and W.J. Marciano, \emph{Higgs decays as a window into the dark sector}, \href{https://doi.org/10.1103/PhysRevD.88.015022} {\emph{Phys. Rev. D} {\bf88} (2013) 015022} [arXiv:1304.4935].
		
		\bibitem{Gopalakrishna2008}
		S. Gopalakrishna, S.Jung, and J.D. Wells, \emph{Higgs boson decays to four fermions through an abelian hidden sector}, \href{https://doi.org/10.1103/PhysRevD.78.055002} {\emph{Phys. Rev. D} {\bf78} (2008) 055002} [arXiv:0801.3456].
		
		\bibitem{Chang2014}
		C.-F. Chang, E. Ma, and T.-C. Yuan, \emph{Multi lepton Higgs Decays through the Dark Portal}, \href{https://doi.org/10.1007/JHEP03(2014)054} {\emph{JHEP} {\bf03} (2014) 054} [arXiv:1308.6071].
		
		\bibitem{Falkowski2014}
		A. Falkowski and R. Vega-Morales, \emph{Exotic Higgs decays in the golden channel}, \href{https://doi.org/10.1007/JHEP12(2014)037} {\emph{JHEP} {\bf12} (2014) 037} [arXiv:1405.1095].
		
		\bibitem{Pospelov2009}
		M. Pospelov, \emph{Secluded U(1) below the weak scale}, \href{https://doi.org/10.1103/PhysRevD.80.095002} {\emph{Phys. Rev. D} {\bf80} (2009) 095002} [arXiv:0811.1030].
		
		\bibitem{Bennett2006}
		G. Bennett \emph{et al.} (Muon $g-2$ Collaboration), \emph{Final Report of the Muon E821 Anomalous Magnetic Moment Measurement at BNL}, \href{https://doi.org/10.1103/PhysRevD.73.072003} {\emph{Phys. Rev. D} {\bf73} (2006) 072003} [arXiv:hep-ex/0602035].
		
		\bibitem{Abi2021}
		B. Abi \emph{et al.} (Muon $g-2$ Collaboration), \emph{Measurement of the Positive Muon Anomalous Magnetic Moment to 0.46 ppm}, \href{https://doi.org/10.1103/PhysRevLett.126.141801} {\emph{Phys. Rev. Lett.} {\bf126} (2021) 141801} [arXiv:2104.03281].
		
		\bibitem{Arkani-Hamed2009}
		N. Arkani-Hamed, D.P. Finkbeiner, T.R. Slatyer, and N. Weiner, \emph{A Theory of Dark Matter}, \href{https://doi.org/10.1103/PhysRevD.79.015014} {\emph{Phys. Rev. D} {\bf79} (2009) 015014} [arXiv:0810.0713].
		
		\bibitem{Pospelov2009a}
		M. Pospelov and A. Ritz, \emph{Astrophysical Signatures of Secluded Dark Matter}, \href{https://doi.org/10.1016/j.physletb.2008.12.012} {\emph{Phys. Lett. B} {\bf671} (2009) 391} [arXiv:0810.1502].
		
		\bibitem{Finkbeiner2007}
		D.P. Finkbeiner and N. Weiner, \emph{Exciting Dark Matter and the INTEGRAL/SPI 511 keV signal}, \href{https://doi.org/10.1103/PhysRevD.76.083519} {\emph{Phys. Rev. D} {\bf76} (2007) 083519} [arXiv:astro-ph/0702587].
		
		\bibitem{Fayet2004}
		P. Fayet, \emph{Light spin 1/2 or spin 0 dark matter particles}, \href{https://doi.org/10.1103/PhysRevD.70.023514} {\emph{Phys. Rev. D} {\bf70} (2004) 023514} [arXiv:hep-ph/0403226].
		
		\bibitem{Curtin2015}
		D. Curtin, R. Essig, S. Gori, and J. Shelton, \emph{Illuminating dark photons with high-energy colliders}, \href{https://doi.org/10.1007/JHEP02(2015)157} {\emph{JHEP} {\bf02} (2015) 157} [arXiv:1412.0018].
		
		\bibitem{Aaij2015}
		R. Aaij \emph{et al.} (The LHCb Collaboration), \emph{Measurement of the forward-backward asymmetry in $Z$/$Y$* → $\mu^{+}\mu^{-}$ decays and determination of the effective weak mixing angle}, \href{https://doi.org/10.1007/JHEP11(2015)190} {\emph{JHEP} {\bf11} (2015) 190} [arXiv:1509.07645]. 
		
		\bibitem{Florian2017}
		D. de Florian, C. Grojean, F. Maltoni, C. Mariotti, A. Nikitenko, M. Pieri, P. Savard, M. Schumacher, and R. Tanaka, \emph{Handbook of LHC Higgs Cross Sections: 4. Deciphering the Nature of the Higgs Sector}, \href{https://doi.org/10.23731/CYRM-2017-002} {\emph{CERN Yellow Reports: Monographs} {\bf2/2017} (2017) CERN–2017–002-M} [arXiv:1610.07922].
		
	\end{thebibliography}
\end{document}